\documentclass[aps,plb,twocolumn,superscriptaddress,preprintnumbers,longbibliography]{revtex4-2}

\usepackage{amsmath}
\usepackage{amssymb}
\usepackage{graphicx}
\usepackage{siunitx}

\usepackage{booktabs}

\usepackage[colorlinks = true, citecolor = blue, linkcolor = blue, urlcolor = blue]{hyperref}
\usepackage{xcolor}
\usepackage{physics}
\usepackage[capitalise]{cleveref} %% has to be loaded after 'amsmath'

\def\bea#1\eea{\begin{align}#1\end{align}}

\newcommand{\bef}{\begin{figure}[h!tb]\centering}
\newcommand{\eef}{\end{figure}}

\allowdisplaybreaks

\begin{document}

\title{Energy--Energy Correlators in $Z$-Tagged Jets in $pp$ Collisions}
\author{Jo\~{a}o Barata}
\email{joao.lourenco.henriques.barata@cern.ch}
\affiliation{CERN, Theoretical Physics Department, CH-1211, Geneva 23, Switzerland}

\author{Zhong-Bo Kang}
\email{zkang@physics.ucla.edu}
\affiliation{Department of Physics and Astronomy, University of California, Los Angeles, CA 90095, USA}
\affiliation{Mani L. Bhaumik Institute for Theoretical Physics, University of California, Los Angeles, CA 90095, USA}
\affiliation{Center for Frontiers in Nuclear Science, Stony Brook University, Stony Brook, NY 11794, USA}

\author{Xo{\'{a}}n Mayo L\'{o}pez}
\email{xoanml@mit.edu}
\affiliation{Center for Theoretical Physics – a Leinweber Institute, Massachusetts Institute of Technology, Cambridge, MA 02139, USA}

\author{Jani Penttala}
\email{janipenttala@physics.ucla.edu}
\affiliation{Department of Physics and Astronomy, University of California, Los Angeles, CA 90095, USA}
\affiliation{Mani L. Bhaumik Institute for Theoretical Physics, University of California, Los Angeles, CA 90095, USA}

\author{Yiyu Zhou}
\email{zyy@impcas.ac.cn}
\affiliation{Institute of Modern Physics, Chinese Academy of Sciences, Lanzhou, Gansu 730000, China}
\affiliation{University of Chinese Academy of Sciences, Beijing 100049, China}

\preprint{CERN-TH-2026-212}
\preprint{MIT-CTP/6102}

\begin{abstract}
We present predictions for the collinear energy-energy correlator (EEC) measured inside \(Z\)-tagged jets in proton-proton collisions across the confinement transition. Our framework combines transverse-momentum-dependent factorization for back-to-back $Z+$jet production with an exclusive differential EEC jet function incorporating perturbative evolution and flavor-dependent non-perturbative fragmentation. The EEC jet function is evaluated at $\mathrm{NLL}^{\prime}$ accuracy, with separate quark and gluon non-perturbative scales constrained by existing $e^+e^-$ and inclusive-jet data. We provide predictions for forward $Z+$jet production at $\sqrt{s}=8~\mathrm{TeV}$  and 13 TeV in the jet transverse-momentum intervals $\pqty{20, 30}$, $\pqty{30, 50}$ and $\pqty{50, 100} \, \mathrm{GeV}$. The EEC distributions as functions of $z_\chi$ exhibit finite plateaus at small $z_\chi$, while the corresponding angular distributions display the characteristic turnover associated with the confinement transition. A flavor decomposition shows that this transition becomes increasingly dominated by quark-initiated jets as the jet transverse momentum increases. Forward \(Z\)-tagged jets therefore provide a clean probe of non-perturbative quark fragmentation and offer a direct test of whether the characteristic confinement scale extracted from \(e^+e^-\) collisions remains universal in a hadronic environment.

\end{abstract}

\maketitle

\section{Introduction} 
Hadronic jets, formed by energetic and collimated cascades of QCD particles produced in high-energy collisions, provide powerful probes of the dynamics of the strong interaction. In this context, a particularly useful class of observables capable of resolving jet substructure is provided by correlators of asymptotic detector operators~\cite{Ore:1979ry,Sveshnikov:1995vi,Tkachov:1995kk,Korchemsky:1999kt,Basham:1978bw,Hofman:2008ar}. The prototypical example is given by correlators of energy flux detectors, \textit{i.e.}, energy correlators. These characterize correlations in the asymptotic energy flows measured by idealized detectors at null infinity, mimicking the role played by experimental detectors; see Ref.~\cite{Moult:2025nhu} for a recent review. The simplest realization of such observables in QCD is the two-point function~\cite{Basham:1979gh,Basham:1978bw,Dixon:2019uzg}, commonly referred to as the Energy-Energy Correlator (EEC):
\begin{align}
\frac{\dd{\Sigma}}{\dd{z_{\chi}}}
=
\int \dd{\sigma_{\mathrm{jet}} \pqty{R}}
\sum_{i \neq j \in J} \frac{p_{Ti} p_{Tj}}{p_{JT}^2}
\delta \pqty{z_\chi - \frac{1-\cos {\chi_{ij}}}{2}}
.
\end{align}
Here the sum runs over pairs of particles inside a jet of radius $R$, $p_{Ti}$ and $p_{Tj}$ denote their transverse momenta, $p_{JT}$ is the transverse momentum of the reconstructed jet, $\chi_{ij}$ is the angular separation of the pair, and $\dd{\sigma_{\mathrm{jet}} \pqty{R}}$ denotes the differential jet cross section.
The measurement angle is set by $\chi = \arccos(1-2z_\chi)$.
Over the past years, the jet EEC has attracted considerable theoretical and experimental interest, both as a precision probe of collinear QCD dynamics~\cite{CMS:2024mlf,Chen:2020vvp,Lee:2022uwt} and for its clean separation of angular regimes dominated by perturbative and non-perturbative physics \cite{Komiske:2022enw,ALICE:2024dfl}.

For sufficiently energetic jets, the collinear EEC distribution exhibits two distinct behaviors.
At larger angular separations, \textit{i.e.}, $\frac{\Lambda_{\mathrm{QCD}}}{p_{JT}} \ll \chi$, the dynamics are determined by perturbative QCD.
Perturbation theory breaks down at scales $\frac{\Lambda_{\mathrm{QCD}}}{p_{JT}} \sim \chi$, and the EEC distribution becomes highly sensitive to hadronization effects. The transition between these regimes has received significant attention recently: its leading non-perturbative corrections have been characterized \cite{Nason:1995np,Korchemsky:1999kt,Dokshitzer:1999sh,
Belitsky:2001ij,Schindler:2023cww,Lee:2024esz}, quantum scaling in the post-confinement regime has been predicted~\cite{Chang:2025kgq}, and the EEC has been related to transverse-momentum-sensitive dihadron fragmentation functions \cite{Lee:2025okn,Kang:2025zto,Guo:2025zwb}. Nonetheless, several aspects of this transition have not been fully understood: for example, the non-perturbative scale characterizing the onset of hadronization has been shown to depend on the flavor content of jets and their intrinsic momentum scale~\cite{Apolinario:2025vtx, Chen:2024quk}, obscuring genuinely universal properties expected to characterize color confinement.

\begin{figure}[t]
\centering
\includegraphics[width = 0.9 \columnwidth]{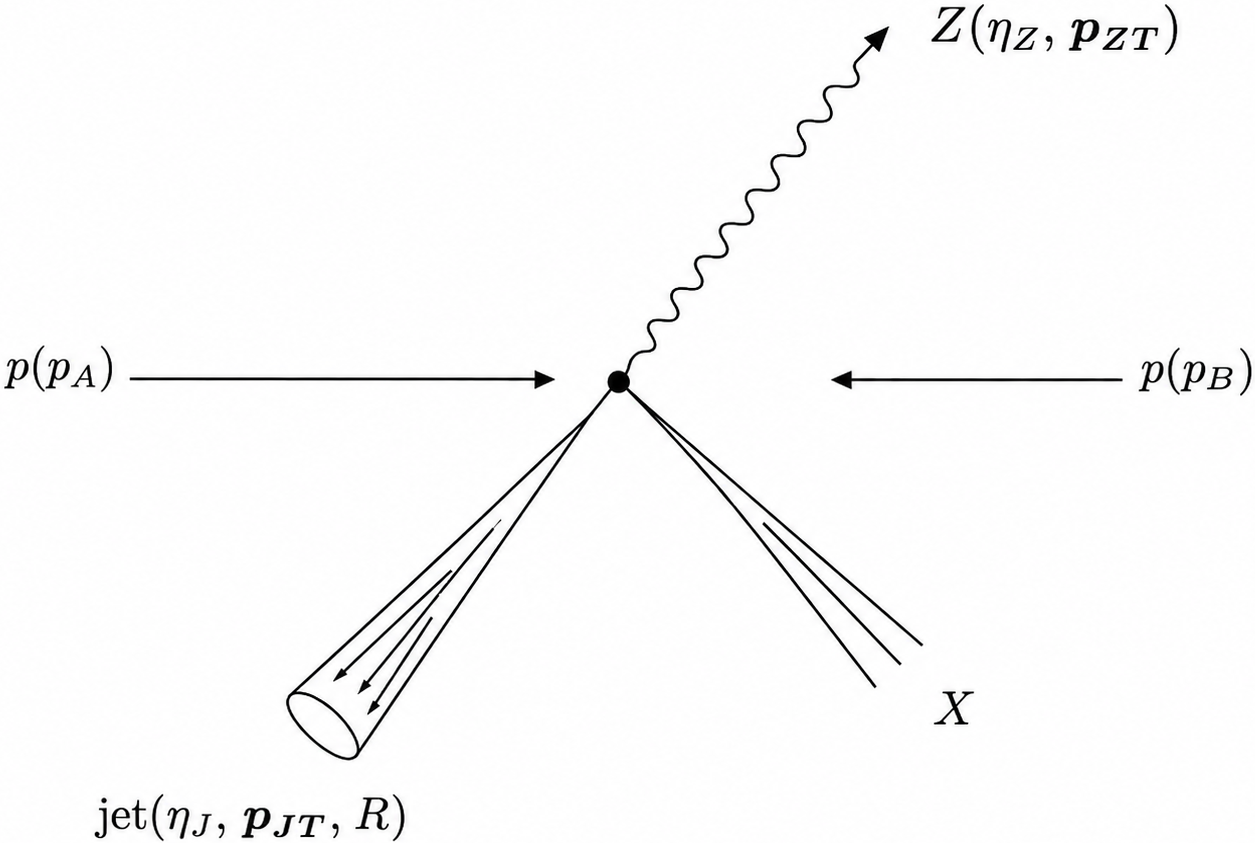}
\caption{
Schematic representation of back-to-back $Z+$jet production in $pp$ collisions, where $X$ denotes unobserved final-state radiation.
}
\label{fig:zjet_schematic}
\end{figure}

A simultaneous and differential description of both the perturbative and non-perturbative regimes is therefore essential to exploit the full information encoded in the EEC. In this work, we propose to use jets produced in association with an electroweak boson, see \cref{fig:zjet_schematic}, to distinguish the quark- and gluon-initiated contributions to the transition scale, providing a direction to extract genuinely  universal properties of the hadronization transition.
In $Z+$jet production, the color-singlet boson provides a clean tag of the underlying hard scattering, while, at forward kinematics, the contributing partonic channels are substantially enriched by quark-initiated jets compared with single-inclusive jet production.
The LHCb Collaboration has measured the longitudinal and transverse distributions of charged hadrons inside forward $Z$-tagged jets in $pp$ collisions at $\sqrt{s}=8~{\rm TeV}$~\cite{LHCb:2019qoc}, followed by multi-differential measurements of identified charged hadrons at $\sqrt{s}=13~\mathrm{TeV}$~\cite{LHCb:2022rky}, demonstrating the experimental accessibility of jet fragmentation in this topology.
From the theory side, factorization frameworks for back-to-back boson-jet production in $pp$ collisions have been developed~\cite{Buffing:2018ggv,Chien:2019gyf} and extended to describe longitudinal and transverse-momentum dependent fragmentation functions inside $Z$-tagged jets~\cite{Kang:2019ahe}.
Very recently, energy correlators measured on the full QCD radiation recoiling against an electroweak boson have also been proposed~\cite{Generet:2026vto}. 
This jet-free construction is complementary to the in-jet EEC considered here, where the measurement is performed inside an identified $Z$-tagged jet and retains sensitivity to the flavor composition of the jet sample.

We focus on this complementary in-jet observable by combining the transverse-momentum-dependent (TMD) factorization of back-to-back $Z+$jet production with the framework of Ref.~\cite{Barata:2024wsu} for describing the EEC across the perturbative-to-non-perturbative transition. 
This extends several recent studies~\cite{Barata:2024wsu,Liu:2024lxy,Herrmann:2025fqy} focused on inclusive jet production for which the quark-gluon composition is not independently controlled and the non-perturbative dynamics are effectively studied through a flavor-averaged input.
For the present application, we retain separate quark- and gluon-dependent non-perturbative inputs, allowing us to study the EEC in a quark-enriched jet sample while preserving its flavor dependence.
This construction provides a unified description of the EEC in exclusive $Z+$jet production across the confinement transition, allowing us to provide predictions for this process in $pp$ collisions at the LHC, for which we show results at $\sqrt{s}=8~\mathrm{TeV}$ and 13 TeV, with the latter presented in Appendix~\ref{App:EEC-at-13-TeV}.

\section{Factorization theorem for EEC in back-to-back $Z+$jet production}
We consider the production of a $Z$ boson recoiling against a jet reconstructed with the anti-$k_T$ algorithm~\cite{Cacciari:2008gp} in $pp$ collisions, as schematically depicted in \cref{fig:zjet_schematic}.
It is convenient to characterize the $Z$-jet system in terms of its transverse-momentum imbalance and average transverse momentum,
\begin{equation}
\boldsymbol{q}_T
\equiv
\boldsymbol{p}_{ZT} + \boldsymbol{p}_{JT},
\qquad
\boldsymbol{p}_T
\equiv
\frac{\boldsymbol{p}_{ZT}-\boldsymbol{p}_{JT}}{2}
.
\end{equation}
The back-to-back region is characterized by the parametric hierarchy $q_T \ll p_T$, with $q_T \equiv \abs{\boldsymbol{q}_T}$ and $p_T \equiv \abs{\boldsymbol{p}_T}$. For the EEC variable $z_{\chi} = \pqty{1-\cos{\chi}}/2$ introduced above, with $\chi \equiv \Delta R
= \sqrt{(\Delta\eta)^2+(\Delta\phi)^2}$ denoting the
angular separation of the particle pair in the
pseudorapidity--azimuth plane, the corresponding transverse-momentum scale is $k_{\chi} = 2\sqrt{z_{\chi}} \, p_{JT}$.
We focus on the collinear EEC regime, in which $k_{\chi}$ is parametrically smaller than the characteristic jet scale $Q_J = p_{JT}R$, \textit{i.e.}, $k_\chi \ll Q_J$. In the combined back-to-back and collinear limits, the EEC measurement modifies the final-state jet sector of the standard $Z+$jet TMD factorization theorem, which can be obtained as a generalization of the photon-tagged jet production factorization theorem~\cite{Buffing:2018ggv}. At leading power, it can therefore be incorporated by replacing the unmeasured exclusive jet function $J_c$ with the exclusive differential EEC jet function $\mathcal{G}^{\mathrm{EEC}}_c$, while leaving the TMD PDFs and the hard and soft functions unmodified. A similar factorization formula for the EEC in
single-inclusive jet production in $pp$ collisions was
presented in Ref.~\cite{Lee:2022uwt}. The resulting factorization formula reads:
\begin{widetext}
% \begin{strip}
\begin{align}
\frac{\dd{\Sigma^{\mathrm{EEC}}}}{\dd{\mathcal{PS}} \dd{z_{\chi}}}
 = &
\sum_{a,b,c} \int \dd{\phi_J}
\int \frac{\dd[2]{\boldsymbol{b}}}{(2\pi)^2} e^{i\boldsymbol{q}_T\cdot\boldsymbol{b}}
f_a \pqty{x_a, b, \mu, \zeta_a}
f_b \pqty{x_b, b, \mu, \zeta_b}
\nonumber \\
& \times
S^{\mathrm{global}}_{n \overline{n} n_J} \pqty{\boldsymbol{b}, \mu}
S^{\mathrm{cs}}_{n_J} \pqty{\boldsymbol{b}, R, \mu}
H_{ab \to cZ} \pqty{p_T, m_Z, \mu}
\,\mathcal{G}^{\mathrm{EEC}}_c \pqty{z_{\chi}, p_{JT} R, \mu}\,.
\label{eq:zjet_eec_factorization}
\end{align}
% \end{strip}
\end{widetext}

\noindent
Here, $\dd{\mathcal{PS}} = \dd{\eta_J} \dd{\eta_Z} \dd{p_T} \dd[2]{\boldsymbol{q}_T}$, $\phi_J$ denotes the jet azimuthal angle, and $\boldsymbol{b}$ is the Fourier conjugate of the transverse-momentum imbalance $\boldsymbol{q}_T$, with $b=|\boldsymbol{b}|$.
The functions $f_a$ and $f_b$ are the TMD PDFs for partons $a$ and $b$ in the incoming protons, depending on their longitudinal momentum fractions $x_{a,b}$, transverse coordinate $b$, the factorization scale $\mu$, and the Collins-Soper rapidity scales $\zeta_{a,b}$.
The global-soft function $S^{\mathrm{global}}_{n \overline{n} n_J}$ describes wide-angle soft radiation associated with the two beam directions $n$, $\overline{n}$ and the measured-jet direction $n_J$, while the collinear-soft function $S^{\mathrm{cs}}_{n_J}$ describes soft radiation collimated along the jet direction and carries the dependence on the jet radius $R$. We use the next-to-leading order (NLO) expressions for the soft functions given in Refs.~\cite{Buffing:2018ggv, Chien:2019gyf}. The hard function $H_{ab \to cZ}$ describes the short-distance production of the $Z$ boson recoiling against the parton $c$ initiating the measured jet.
We include the $q \overline{q} \to Zg$ and $qg \to Zq$ channels through the hard functions evaluated at NLO~\cite{Becher:2011fc, Arnold:1988dp}. It is instructive to note that this factorized formula follows the implementation of $Z+$jet TMD factorization in Ref.~\cite{Kang:2019ahe}, but adopts the convention of the TMD Handbook~\cite{Boussarie:2023izj}, in which the relevant part of the global soft function is absorbed into the unsubtracted TMD PDFs, thereby defining subtracted TMD PDFs that are free of rapidity divergences.

We consider the EEC distribution obtained by integrating \cref{eq:zjet_eec_factorization} over a given phase-space region while keeping $z_\chi$ differential, and normalize it to the corresponding $Z+$jet cross section $\sigma_{ZJ}$ evaluated over the same kinematic region. The normalized EEC distribution is defined as:
\begin{equation}
F_{\mathrm{EEC}} \pqty{z_{\chi}}
\equiv
\frac{1}{\sigma_{ZJ}}
\frac{\dd{\Sigma^{\mathrm{EEC}}}}{\dd{z_{\chi}}}
. \label{eq:normalized_eec}
\end{equation}
The final-state parton index $c$ in \cref{eq:zjet_eec_factorization} allows us to further decompose the prediction into quark- and gluon-initiated contributions,
\begin{equation}
F_{\mathrm{EEC}} \pqty{z_{\chi}}
=
F_q \pqty{z_{\chi}}
+
F_g \pqty{z_{\chi}}
,
\end{equation}
where $F_q$ includes both quark and antiquark channels.
Both $F_q$ and $F_g$ are normalized to the same $Z+$jet cross section $\sigma_{ZJ}$.

\section{Exclusive differential EEC jet function}  
We construct the exclusive differential EEC jet function $\mathcal{G}^{\mathrm{EEC}}_i$ entering \cref{eq:zjet_eec_factorization} from the inclusive differential EEC jet function $j_i^{\mathrm{EEC}}$ in coordinate space.
To distinguish the transverse coordinate associated with the EEC measurement from that conjugate to the $Z$-jet recoil, we denote the former by $b_{\chi}$, conjugate to $k_{\chi}$.
At the jet scale $\mu_J$, the two functions are related by:
\begin{align}\label{eq:eec_matching}
\mathcal{G}^{\mathrm{EEC}}_i \pqty{b_{\chi}, Q_J, \mu_J}
& =
\sum_j \int_0^1 \dd{x} x^2 \mathcal{J}_{ij} \pqty{x, Q_J, \mu_J}
\\ \notag
& \qquad \qquad \times
j_j^{\mathrm{EEC}} \pqty{\frac{b_{\chi}}{x},\mu_J}
,
\end{align}
where the matching coefficients $\mathcal{J}_{ij}$ are the same short-distance coefficients that describe the perturbative matching of a fragmenting jet function onto the standard collinear fragmentation functions~\cite{Kang:2019ahe}.
We include them at one loop using the anti-$k_T$ results of Refs.~\cite{Chien:2015ctp,Kang:2023elg}. The same matching
structure was found for the EEC in single-inclusive jet
production in Ref.~\cite{Lee:2022uwt}.

Following Refs.~\cite{Barata:2024wsu,Herrmann:2025fqy}, non-perturbative fragmentation dynamics are introduced through a boundary condition for $j_i^{\mathrm{EEC}}$ at its natural $b_{\chi}$-dependent scale.
At the accuracy considered here, we take:
\begin{equation}
j_i^{\mathrm{EEC}} \pqty{b_{\chi}, \mu_{b_{\chi}}}
=
\bqty{
1 + \frac{\alpha_s \pqty{\mu_{b_{\chi}}}}{4 \pi}
j_{1,i}
}
e^{-a_i b_{\chi}}
, \label{eq:eec_initial_condition}
\end{equation}
where $i=q$, $g$, and $j_{1,i}$ are the NLO EEC jet-function constants.
At this order, the constants entering the differential EEC jet function coincide with those of the cumulative EEC jet functions computed in Ref.~\cite{Dixon:2019uzg}.
The canonical coordinate-space scale is regulated using the standard $b^*$ prescription,
\begin{equation}
\mu_{b_{\chi}}
=
\frac{2 e^{-\gamma_E}}{b_\chi^*}
,
\quad
b_{\chi}^*
\equiv
\frac{b_{\chi}}
{\sqrt{1 + b_{\chi}^2/b_{\max}^2}}
, \label{eq:bstar}
\end{equation}
with $b_{\max} \equiv 2e^{-\gamma_E} \, \mathrm{GeV}^{-1}$ and $\gamma_E$ being the Euler-Mascheroni constant, such that $\mu_{b_{\chi}} \to 1~\mathrm{GeV}$ at large $b_\chi$.
Non-perturbative transverse-momentum effects associated with fragmentation are encoded in the exponential factors $e^{-a_i b_{\chi}}$, with $a_i$ allowed to depend on the flavor of the initiating parton.
Motivated by fits to quark-dominated $e^+e^-$ and gluon-dominated $pp$ data in Refs.~\cite{Herrmann:2025fqy, Barata:2024wsu}, we take:
\begin{equation}
a_q = 2.31~\mathrm{GeV},
\quad
a_g = 3.8~\mathrm{GeV}
. \label{eq:NP_parameters}
\end{equation}

Starting from \cref{eq:eec_initial_condition}, the inclusive differential EEC jet function is evolved to the characteristic jet scale $\mu_J \sim Q_J$ through its timelike renormalization group (RG) equation:
\begin{equation}
\frac{\dd{}}{\dd{\ln{\mu^2}}}
j_i^{\mathrm{EEC}} \pqty{b_{\chi}, \mu}
=
\sum_j \int_0^1 \dd{y}
y^2 j_j^{\mathrm{EEC}} \pqty{\frac{b_{\chi}}{y}, \mu}
P^T_{ji} \pqty{y, \mu}
, \label{eq:inclusive_eec_rge}
\end{equation}
where $P^{T}_{ji}$ denote the timelike splitting functions, included through NLO \cite{Curci:1980uw,Furmanski:1980cm}, and the sum runs over quark and gluon daughter partons.
Since the natural scale $\mu_{b_\chi}$ depends on $b_\chi$, \cref{eq:eec_initial_condition} defines the boundary condition along the curve $\mu = \mu_{b_{\chi}} \pqty{b_{\chi}}$.
Moreover, the convolution in \cref{eq:inclusive_eec_rge} couples different values of $b_\chi$ through $j_j^{\mathrm{EEC}} \pqty{\frac{b_{\chi}}{y}, \mu}$.
The resulting coupled system must therefore be evolved from this moving boundary to the common jet scale $\mu_J$.
Details of the numerical solution of this evolution equation and its extension to higher perturbative accuracy will be presented separately in Ref.~\cite{EECJetFunction:toappear}.

The momentum-space exclusive EEC jet function at the matching scale is then obtained through the Hankel transform:
\begin{align}
& \mathcal{G}^{\mathrm{EEC}}_i \pqty{z_{\chi}, Q_J, \mu_J}
=
2 p_{JT}^2
\nonumber \\
& \times
\int_0^{\infty} \dd{b_{\chi}}
b_{\chi}
J_0 \pqty{2 \sqrt{z_{\chi}} \, p_{JT} b_{\chi}}
\mathcal{G}^{\mathrm{EEC}}_i \pqty{b_{\chi}, Q_J, \mu_J}
. \label{eq:eec_bessel}
\end{align}
The exponential factor in \cref{eq:eec_initial_condition} suppresses the large-$b_{\chi}$ region, and this suppression is inherited by the evolved and matched jet function in coordinate space.
Consequently, the Hankel transform remains finite as $z_{\chi} \to 0$.
In particular,
\begin{align}
& \lim_{z_{\chi} \to 0}
\mathcal{G}^{\mathrm{EEC}}_i \pqty{z_{\chi}, Q_J, \mu_J}
=
2 p_{JT}^2
\nonumber \\
& \qquad \qquad \qquad \times
\int_0^{\infty} \dd{b_{\chi}}
b_{\chi}
\mathcal{G}^{\mathrm{EEC}}_i \pqty{b_{\chi}, Q_J, \mu_J}
, \label{eq:eec_plateau}
\end{align}
which gives rise to the finite small-$z_{\chi}$ plateau discussed in the following section.
The corresponding small-angle behavior has been observed in in-jet EEC measurements in $pp$ collisions and event-wide EEC measurement in $e^+e^-$ collisions \cite{CMS:2024mlf, ALICE:2024dfl, OPAL:1993pnw, Electron-PositronAlliance:2025fhk}.

RG consistency requires the exclusive differential EEC jet function to obey the same multiplicative RG equation as the unmeasured exclusive jet function $J_i$ entering the production cross section $\sigma_{ZJ}$~\cite{Kang:2019ahe},
\begin{equation}
\mu\frac{\dd{}}{\dd{\mu}}
\mathcal{G}^{\mathrm{EEC}}_i \pqty{z_{\chi}, Q_J, \mu}
=
\gamma_J^i \pqty{\mu}
\mathcal{G}^{\mathrm{EEC}}_i \pqty{z_{\chi}, Q_J, \mu}
, \label{eq:exclusive_eec_rge}
\end{equation}
where the jet anomalous dimension $\gamma_J^i$ is given by:
\begin{equation}
\gamma_J^i \pqty{\mu}
=
-2 \Gamma_{\mathrm{cusp}}^i \pqty{\alpha_s}
\ln(\frac{Q_J}{\mu})
+
\gamma^i \pqty{\alpha_s}
, \label{eq:jet_anomalous_dimension}
\end{equation}
with $\Gamma_{\mathrm{cusp}}^i$ and $\gamma^i$ being the cusp and non-cusp anomalous dimensions, which we include through NLO and LO respectively \cite{Becher:2009th, Jain:2011xz, Liu:2012sz}.
Consequently, after constructing $\mathcal{G}^{\mathrm{EEC}}_i$ at $\mu_J$, it is evolved to the hard scale $\mu_H$ entering \cref{eq:zjet_eec_factorization} as:
\begin{align}
\mathcal{G}^{\mathrm{EEC}}_i \pqty{z_{\chi}, Q_J, \mu_H}
& =
\mathcal{G}^{\mathrm{EEC}}_i \pqty{z_{\chi}, Q_J, \mu_J}
\nonumber \\
& \quad \times
\exp[
\int_{\mu_J}^{\mu_H}
\frac{\dd{\mu'}}{\mu'}
\gamma_J^i \left(\mu'\right)
]. 
\label{eq:eec_hard_evolution}
\end{align}
With the perturbative ingredients specified above and the two-loop running of $\alpha_s$, we evaluate the exclusive differential EEC jet function at NLL$'$ accuracy.
To consistently account for heavy-flavor thresholds throughout this construction and its RG evolution, we employ a variable-flavor-number scheme, changing $n_f$ across heavy-quark thresholds and matching $\alpha_s$ accordingly.
The resulting $n_f \pqty{\mu}$ is used consistently in the perturbative boundary condition, timelike splitting kernels, exclusive matching coefficients, and jet anomalous dimension.
We impose continuity of the differential EEC jet function across heavy-quark thresholds.
Further details on the perturbative evolution and the treatment of heavy flavors will be presented in Ref.~\cite{EECJetFunction:toappear}.

\section{Phenomenological implementation} 
The phenomenological implementation of \cref{eq:normalized_eec} presented below focuses on $pp$ collisions at $\sqrt{s} = 8~\mathrm{TeV}$ and anti-$k_T$ jets with radius $R=0.5$.
We impose the following kinematic cuts: $2.5 < \eta_J < 4.0$, $2.0 < \eta_Z < 4.5$, and
$7 \pi/8 < \abs{\Delta \phi_{ZJ}} < \pi$, and consider the three jet transverse-momentum intervals $20~\mathrm{GeV} < p_{JT} < 30~\mathrm{GeV}$,
$30~\mathrm{GeV} < p_{JT} < 50~\mathrm{GeV}$, and $50~\mathrm{GeV} < p_{JT} < 100~\mathrm{GeV}$.
These choices are motivated by the kinematic coverage of the LHCb measurements of fragmentation inside forward $Z$-tagged jets~\cite{LHCb:2019qoc}.
Corresponding predictions for $\sqrt{s} = 13~\mathrm{TeV}$, obtained with the same kinematic cuts and theoretical setup, are presented in Appendix~\ref{App:EEC-at-13-TeV}.
The central hard and jet scales are chosen as:
\begin{equation}
\mu_H = \sqrt{p_T^2 + m_Z^2},
\quad
\mu_J = Q_J = p_{JT}R
.
\end{equation}
The $Z$-boson transverse momentum is integrated over $0~\mathrm{GeV} < p_{ZT} < 100~\mathrm{GeV}$.
We use CT14NLO collinear PDFs \cite{Dulat:2015mca} as input and otherwise follow Ref.~\cite{Kang:2019ahe} for the TMD evolution, soft functions, and non-perturbative initial-state recoil model.
Distributions differential in the opening angle $\chi$ are obtained from \cref{eq:normalized_eec} using the Jacobian:
\begin{equation}
\frac{\dd{\Sigma^{\mathrm{EEC}}}}{\dd{\chi}}
=
\frac{\sin{\chi}}{2}
\frac{\dd{\Sigma^{\mathrm{EEC}}}}{\dd{z_{\chi}}}
,
\end{equation}
with the corresponding normalized distribution being therefore:
\begin{equation}
F_{\mathrm{EEC}} \pqty{\chi}
\equiv
\frac{1}{\sigma_{ZJ}}
\frac{\dd{\Sigma^{\mathrm{EEC}}}}{\dd{\chi}}
=
\frac{\sin{\chi}}{2}
F_{\mathrm{EEC}} \pqty{z_{\chi} \pqty{\chi}}
. \label{eq:EEC_differential_chi}
\end{equation}
For each $p_{JT}$ interval, we define the characteristic jet momentum $\expval{p_{JT}}$ as the cross-section-weighted average:
\begin{equation}\label{eq:average-P_TJ-definition}
\expval{p_{JT}}
=
\frac{
\displaystyle
\int_{\mathrm{bin}} \dd{\mathcal{PS}}
p_{JT}
\frac{\dd{\sigma_{ZJ}}}{\dd{\mathcal{PS}}}
}
{
\displaystyle
\int_{\mathrm{bin}} \dd{\mathcal{PS}}
\frac{\dd{\sigma_{ZJ}}}{\dd{\mathcal{PS}}}
}.
\end{equation}

\begin{figure}[!t]
\centering
\includegraphics[width=0.99\columnwidth]{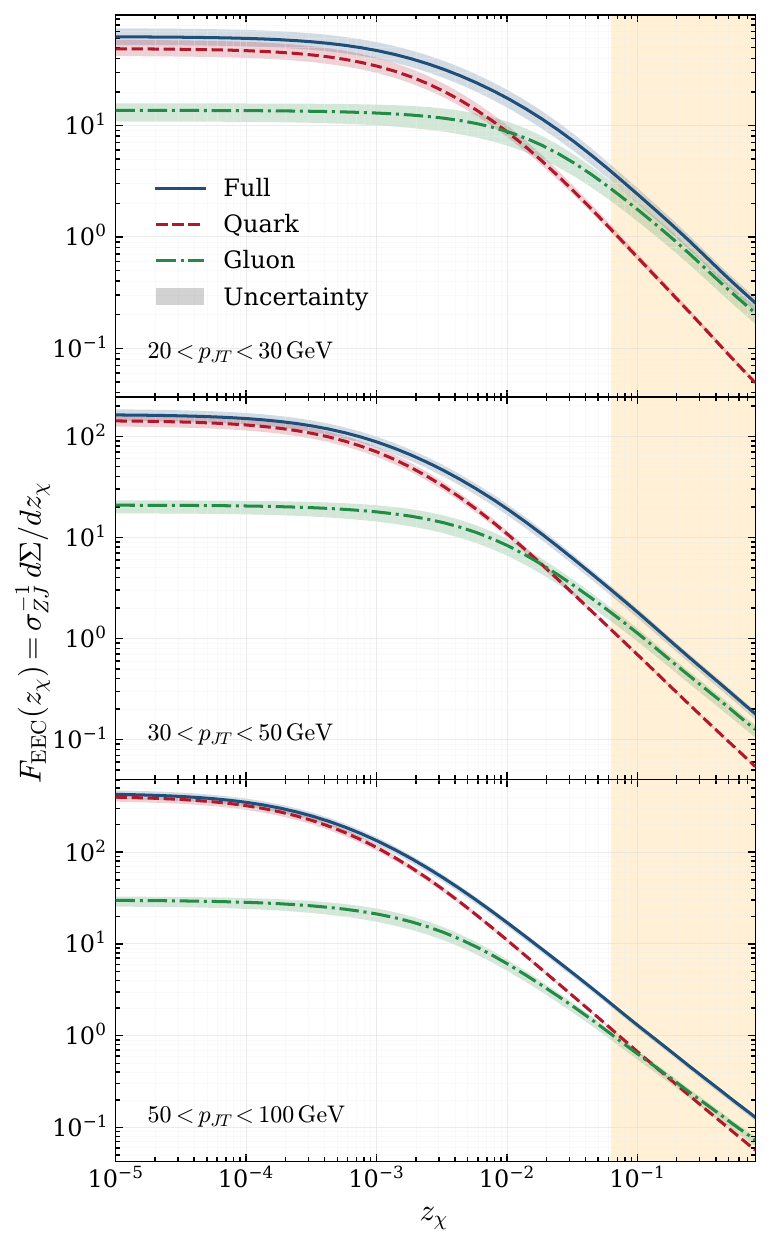}
\caption{Normalized EEC distribution $F_{\rm EEC}(z_\chi)$, \cref{eq:normalized_eec}, at $\sqrt{s}=8~\mathrm{TeV}$ as a function of $z_\chi$ for the three $p_{JT}$ bins considered.
The solid blue curve shows the full prediction, while dashed red and dash-dotted green curves show the quark- and gluon-initiated contributions $F_q \pqty{z_{\chi}}$ and $F_g \pqty{z_{\chi}}$ respectively.
The corresponding bands indicate the theoretical uncertainty associated with resummation scale variation, obtained by independently varying $\mu_J$ in \cref{eq:eec_matching} and $\mu_H$ in \cref{eq:zjet_eec_factorization} by factors of 2 and taking the envelope of the resulting predictions.
The orange region marks $k_{\chi} \geq Q_J$, where the collinear hierarchy underlying the derivation of \cref{eq:zjet_eec_factorization} is no longer parametrically satisfied and the factorization formula is not expected to be reliable.
For $R=0.5$, the boundary $k_{\chi} = Q_J$ corresponds to $z_{\chi} = 0.063$.}
\label{fig:F-vs-z_chi}
\end{figure}

The normalized EEC distribution $F_{\mathrm{EEC}} \pqty{z_{\chi}}$ for the three jet transverse-momentum intervals introduced above is shown in \cref{fig:F-vs-z_chi}, together with its quark- and gluon-initiated components, $F_q \pqty{z_{\chi}}$ and $F_g \pqty{z_{\chi}}$ respectively.
The bands indicate the theoretical uncertainty from resummation-scale variations, obtained by independently varying the matching scale $\mu_J$ in \cref{eq:eec_matching} and the hard scale $\mu_H$ in \cref{eq:zjet_eec_factorization} by factors of 2 and taking the envelope of the resulting predictions.
They therefore quantify the residual factorization-scale dependence of the calculation, providing an estimate of missing higher-order corrections, but do not include uncertainties associated with the non-perturbative parameters $a_q$ and $a_g$.
The orange-shaded region indicates $k_{\chi} \geq Q_J$, where the collinear hierarchy is no longer parametrically satisfied and \cref{eq:zjet_eec_factorization} is not expected to be reliable. 

\begin{figure}[!t]
\centering
\includegraphics[width=0.99\columnwidth]{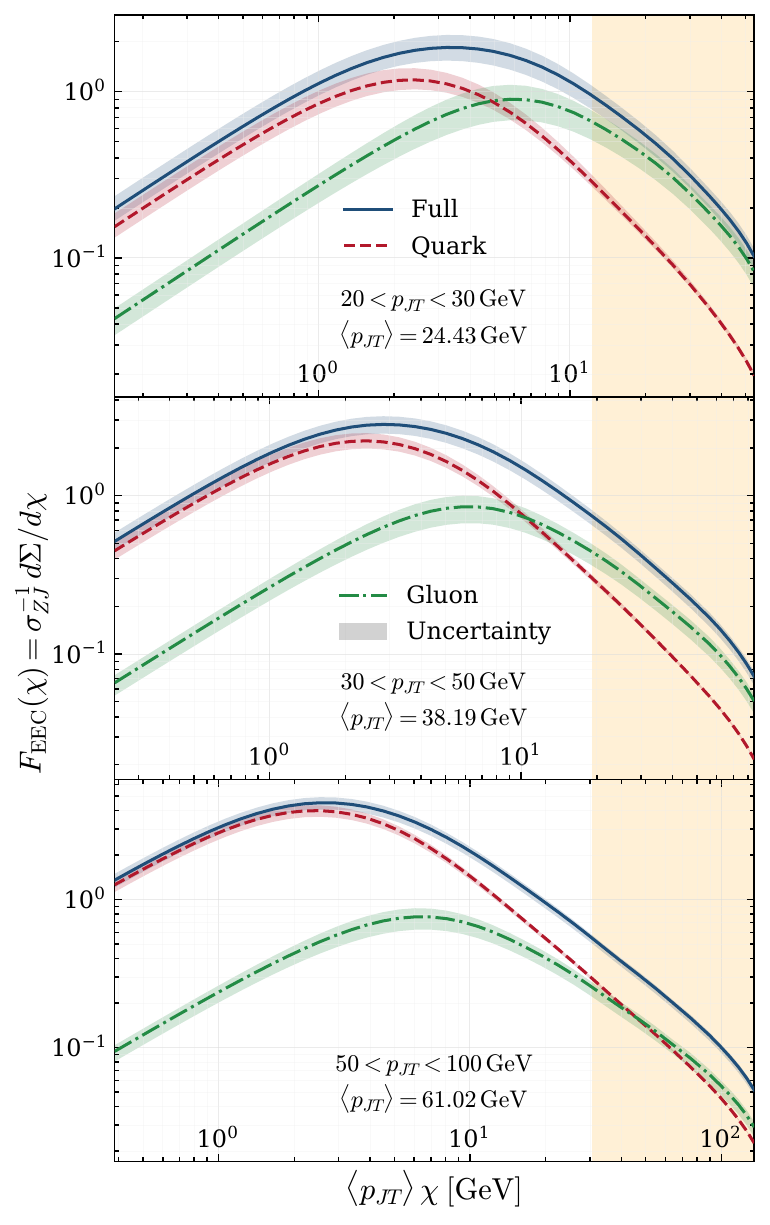}
\caption{
Normalized EEC distribution $F_{\mathrm{EEC}} \pqty{\chi}$, \cref{eq:EEC_differential_chi}, at $\sqrt{s}=8~\mathrm{TeV}$ as a function of the scaled angle $\expval{p_{JT}} \chi$ for the three $p_{JT}$ bins considered.
The bin average jet transverse momentum is defined in \cref{eq:average-P_TJ-definition}.
The solid blue curve shows the full prediction, while dashed red and dash-dotted green curves show the quark- and gluon-initiated contributions $F_q \pqty{\chi}$ and $F_g \pqty{\chi}$ respectively.
The corresponding bands indicate the theoretical uncertainty associated with resummation scale variation, obtained by independently varying $\mu_J$ in \cref{eq:eec_matching} and $\mu_H$ in \cref{eq:zjet_eec_factorization} by factors of 2 and taking the envelope of the resulting predictions.
The orange region marks $k_{\chi} \geq Q_J$, where the collinear hierarchy underlying the derivation of \cref{eq:zjet_eec_factorization} is no longer parametrically satisfied and the factorization formula is not expected to be reliable.
For $R=0.5$, the boundary $k_{\chi} = Q_J$ corresponds to $\chi = 0.505$.
}
\label{fig:F-vs-chi}
\end{figure}

At small $z_{\chi}$, the EEC distributions approach the finite plateau identified in \cref{eq:eec_plateau}.
The flavor decomposition shows that, in the forward $Z+$jet kinematics considered here, this non-perturbative regime is dominated by the quark-initiated contribution.
This dominance becomes increasingly pronounced with increasing jet transverse momentum, reflecting the changing flavor composition of the $Z$-tagged jet sample across the three $p_{JT}$ bins.
At the same time, increasing $p_{JT}$ probes progressively more perturbative fragmentation dynamics at fixed $z_{\chi}$, since the characteristic transverse scale $k_{\chi}$ itself increases.
This does not extend the range of validity in $z_{\chi}$ of the collinear hierarchy underlying \cref{eq:zjet_eec_factorization}, since $k_{\chi} / Q_J = 2\sqrt{z_\chi} / R$ is independent of $p_{JT}$.
Rather, increasing the jet transverse momentum shifts the transition between perturbative and non-perturbative fragmentation toward smaller $z_{\chi}$, therefore enlarging the perturbatively dominated region within the domain of validity of collinear factorization.

The same predictions are shown in \cref{fig:F-vs-chi} as distributions in the opening angle $\chi$, plotted as a function of the rescaled variable $\expval{p_{JT}} \chi$.
In this representation, the small-$z_{\chi}$ plateau is mapped by the Jacobian $\dd{z_{\chi}} / \dd{\chi} = \sin(\chi) / 2$ into a distribution that vanishes linearly as $\chi \to 0$, giving rise to the characteristic turnover of the EEC across the confinement transition.
In the collinear limit, the small-angle approximation yields $k_\chi = 2p_{JT} \sin(\chi / 2) \simeq p_{JT}\chi$, so that $\expval{p_{JT}} \chi$ provides a convenient proxy for the characteristic transverse scale probed by the correlator in each $p_{JT}$ bin.
The flavor decomposition exhibits the same trend already observed in the $z_\chi$ distributions, \textit{i.e.}, increasing quark dominance with increasing $p_{JT}$.
Consequently, at larger jet transverse momentum, the confinement transition of the full EEC distribution is increasingly controlled by the transition scale of the quark contribution.
In this presentation the rescaled variable $\expval{p_{JT}} \chi$ makes the separation between the quark and gluon transition scales more explicit: the quark contribution turns over at a smaller value of $\expval{p_{JT}} \chi$ than the gluon contribution, reflecting the hierarchy $a_q < a_g$.
Taken together, these features show that forward $Z+$jet production in $pp$ collisions provides increasingly clean access to the quark confinement transition as the jet transverse momentum increases.

An important aspect of the presented prediction is that the quark non-perturbative parameter is constrained independently of the $Z+$jet process.
The value in \cref{eq:NP_parameters} used in this work corresponds to the characteristic scale extracted from a global analysis of EEC measurements in $e^+e^-$ collisions, which are dominated by quark-initiated jets \cite{Herrmann:2025fqy}.
Since the non-perturbative contribution to the collinear EEC originates from final-state hadronization, the underlying non-perturbative quark dynamics are expected to be process independent.
A measurement of the EEC in the quark-enriched $Z+$jet topology would therefore test whether the characteristic scale extracted from $e^+e^-$ collisions also describes the confinement transition in a hadronic collision environment.

The gluon non-perturbative input is less precisely constrained.
The value in \cref{eq:NP_parameters} used in these results was extracted from the fit to CMS single-inclusive jet EEC data performed in Ref.~\cite{Barata:2024wsu}.
That analysis was performed at LL accuracy using a single flavor-independent non-perturbative parameter, despite the expectation that quark and gluon fragmentation should be characterized by different non-perturbative scales.
Since the inclusive-jet sample is predominantly gluon initiated in the kinematic region relevant to that extraction, the fitted value provides a natural estimate for the gluon non-perturbative scale, given that no flavor-separated determination of $a_g$ is available yet.
In the forward $Z+$jet process considered here, however, the gluon contribution is subleading and becomes progressively smaller relative to the quark contribution as $p_{JT}$ increases, which limits the sensitivity of the full prediction to the precise value of $a_g$.
A dedicated higher-accuracy fit to inclusive-jet EEC data with independent quark and gluon non-perturbative inputs would enable a more precise determination of $a_g$ and further sharpen the predictions presented here. Ultimately, a global analysis of EEC measurements in $e^+e^-$ collisions, $Z$-tagged jets, and single-inclusive jet production would enable a systematic test of the universality and flavor dependence of the non-perturbative dynamics governing the perturbative-to-non-perturbative transition in the EEC.

\section{Conclusions} 
We have presented theoretical predictions for the collinear EEC in $Z$-tagged jets in $pp$ collisions across the confinement transition.
Our framework combines the TMD factorization for back-to-back $Z+$jet production with a description of the EEC that incorporates both perturbative evolution and non-perturbative fragmentation effects.
The EEC measurement is encoded in the exclusive differential EEC jet function $\mathcal{G}^{\mathrm{EEC}}$, constructed by evolving the inclusive differential EEC jet function from its natural coordinate-space scale to the jet scale and subsequently matching onto its exclusive counterpart.
Non-perturbative fragmentation effects enter through flavor-dependent boundary conditions for the inclusive differential jet function, parametrized by the quark and gluon scales $a_q$ and $a_g$.

We have investigated the flavor dependence of the confinement transition using both the $z_{\chi}$ and angular representations of the EEC for the jet transverse-momentum intervals $\pqty{20, 30}$, $\pqty{30, 50}$ and $\pqty{50, 100} \, \mathrm{GeV}$ at both $\sqrt{s}=8$ and 13 TeV.
The decomposition into quark- and gluon-initiated contributions shows that, in the forward $Z+$jet kinematics considered here, the confinement transition becomes increasingly quark dominated as $p_{JT}$ increases.
At fixed $p_{JT}$, increasing the center-of-mass energy from $\sqrt{s}=8$ to 13 TeV further enhances the quark enrichment of the jet sample and, correspondingly, the quark dominance of the EEC.
Consequently, forward $Z+$jet production provides increasingly clean access to the non-perturbative quark fragmentation dynamics governing the EEC transition.
A measurement of this observable would therefore offer a direct test of whether the characteristic quark non-perturbative transition scale inferred from $e^+e^-$ data also describes the confinement transition in a hadronic collision environment.

More generally, these results identify $Z$-tagged jets as a promising laboratory for resolving the flavor dependence of the transition between perturbative and non-perturbative regions in energy correlators.
A dedicated higher-accuracy analysis of inclusive-jet EEC data with independent quark and gluon non-perturbative inputs would provide a more precise determination of the gluon scale and further sharpen the predictions presented here.
Together with future measurements of the EEC in flavor-sensitive jet samples, including jets produced in
deep-inelastic scattering at the future Electron-Ion
Collider, this would enable a systematic test of the universality and flavor dependence of the non-perturbative dynamics governing the confinement transition.

\section*{Acknowledgments} 
The authors thank Xiaoyuan Zhang and Congyue Zhang for valuable discussions during the development of this work. Z.K. and J.P. are supported by the National Science Foundation under grant No.~PHY-2515057, and by the U.S. Department of Energy, Office of Science, Office of Nuclear Physics, within the framework of the Saturated Glue (SURGE) Topical Theory Collaboration. This material is based upon work supported by the U.S. Department of Energy, Office of Science, Office of Nuclear Physics under grant Contract Number DE-SC0011090.

\appendix
\section{EEC in exclusive $Z+$jet production at $\sqrt{s}=13~\mathrm{TeV}$ in $pp$ collisions} \label{App:EEC-at-13-TeV}

\begin{figure}[!t]
\centering
\includegraphics[width=0.99\columnwidth]{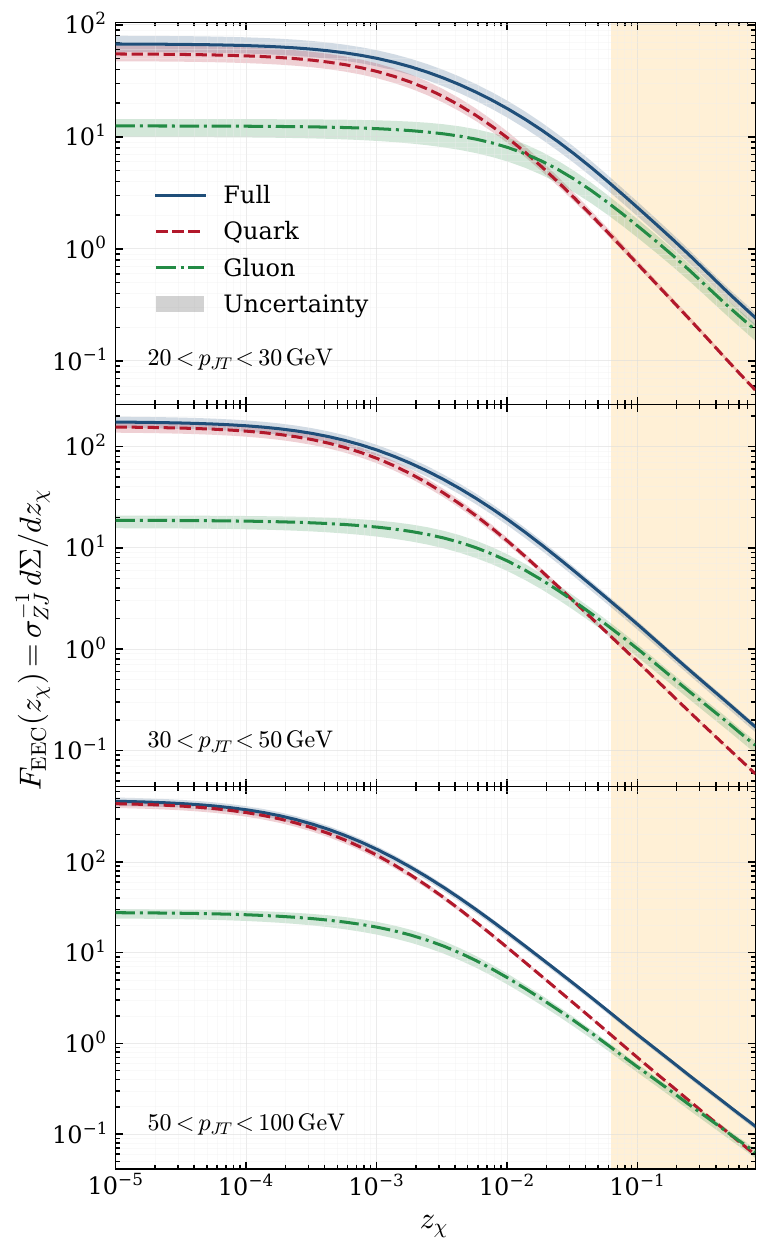}
\caption{
Normalized EEC distribution $F_{\mathrm{EEC}} \pqty{z_{\chi}}$, \cref{eq:normalized_eec}, at $\sqrt{s}=13~\mathrm{TeV}$ as a function of $z_\chi$ for the three $p_{JT}$ bins considered.
The solid blue curve shows the full prediction, while dashed red and dash-dotted green curves show the quark- and gluon-initiated contributions  $F_q \pqty{z_{\chi}}$ and $F_g \pqty{z_{\chi}}$ respectively. 
The corresponding bands indicate the theoretical uncertainty associated with resummation scale variation, obtained by independently varying $\mu_J$ in \cref{eq:eec_matching} and $\mu_H$ in \cref{eq:zjet_eec_factorization} by factors of 2 and taking the envelope of the resulting predictions.
The orange region marks $k_\chi\geq Q_J$, where the collinear hierarchy underlying the derivation of \cref{eq:zjet_eec_factorization} is no longer parametrically satisfied and the factorization formula is not expected to be reliable.
For $R=0.5$, the boundary $k_{\chi} = Q_J$ corresponds to $z_{\chi} = 0.063$.}
\label{fig:F-vs-z_chi-13TeV}
\end{figure}
\begin{figure}[!t]
\centering
\includegraphics[width=0.99\columnwidth]{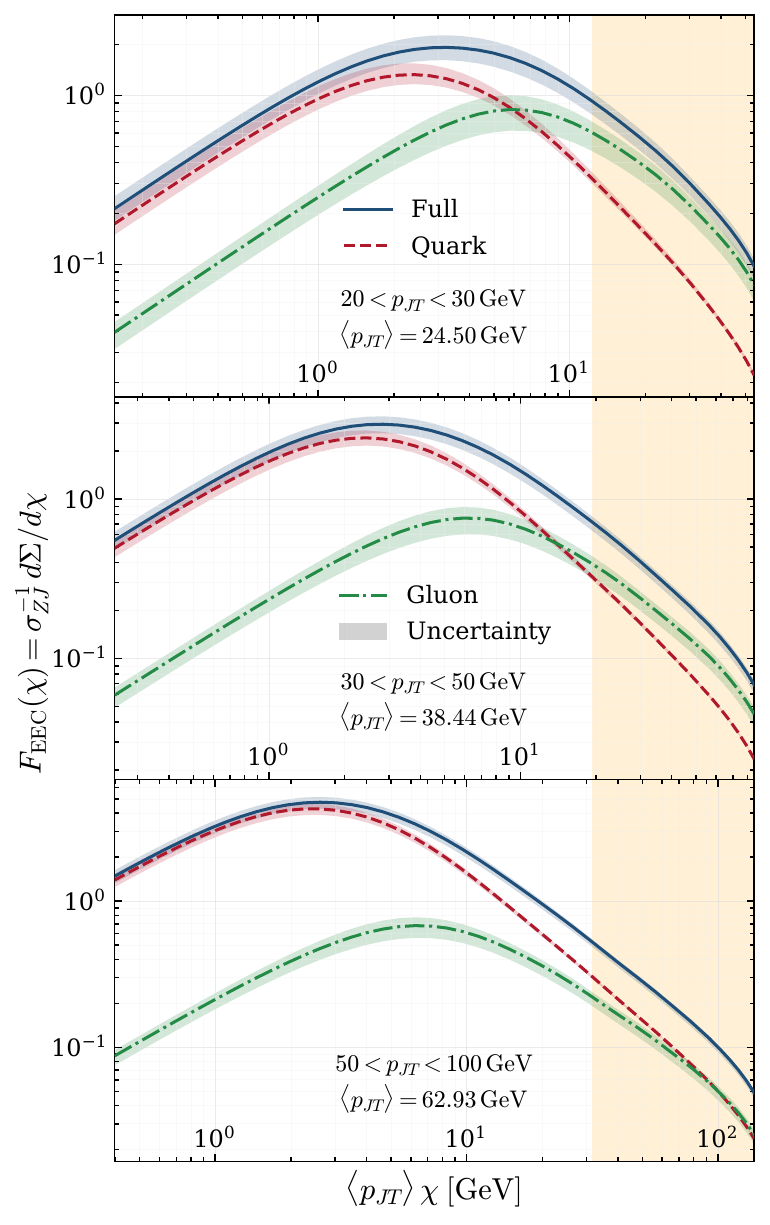}
\caption{
Normalized EEC distribution $F_{\mathrm{EEC}} \pqty{\chi}$, \cref{eq:EEC_differential_chi}, at $\sqrt{s}=13~\mathrm{TeV}$ as a function of the scaled angle $\expval{p_{JT}} \chi$ for the three $p_{JT}$ bins considered.
The bin average jet transverse momentum is defined in \cref{eq:average-P_TJ-definition}.
The solid blue curve shows the full prediction, while dashed red and dash-dotted green curves show the quark- and gluon-initiated contributions $F_q \pqty{\chi}$ and $F_g \pqty{\chi}$ respectively.
The corresponding bands indicate the theoretical uncertainty associated with resummation scale variation, obtained by independently varying $\mu_J$ in \cref{eq:eec_matching} and $\mu_H$ in \cref{eq:zjet_eec_factorization} by factors of 2 and taking the envelope of the resulting predictions.
The orange region marks $k_{\chi} \geq Q_J$, where the collinear hierarchy underlying the derivation of \cref{eq:zjet_eec_factorization} is no longer parametrically satisfied and the factorization formula is not expected to be reliable.
For $R=0.5$, the boundary $k_{\chi} = Q_J$ corresponds to $\chi = 0.505$.} 
\label{fig:F-vs-chi-13TeV}
\end{figure}

In this Appendix, we present predictions for the collinear EEC in exclusive $Z+$jet production in $pp$ collisions at $\sqrt{s}=13~\mathrm{TeV}$, corresponding to the center-of-mass energy of more recent measurements of $Z+$jet production and $Z$-tagged jet observables by the LHCb, CMS and ATLAS Collaborations \cite{CMS:2021iwu,LHCb:2022rky,ATLAS:2022nrp}.
The normalized EEC distributions $F_{\mathrm{EEC}}$ for the same three jet transverse-momentum intervals considered in the main text are shown in \cref{fig:F-vs-z_chi-13TeV,fig:F-vs-chi-13TeV} as functions of $z_{\chi}$ and $\chi$ respectively.
As in the main text, the full predictions are shown together with their quark- and gluon-initiated contributions and the corresponding bands indicating the theoretical uncertainty associated with residual resummation-scale dependence.
The orange-shaded regions mark $k_{\chi} \geq Q_J$, where the collinear hierarchy underlying \cref{eq:zjet_eec_factorization} is no longer satisfied and the factorization formula is not expected to be reliable.

The final-state jet sector, both for the unmeasured and EEC-measured jet functions, is independent of the hadronic center-of-mass energy at fixed jet transverse momentum.
The dependence of \cref{eq:zjet_eec_factorization} on $\sqrt{s}$ enters primarily through the longitudinal momentum fractions $x_a$ and $x_b$ probed in the TMD PDFs.
Changing $\sqrt{s}$ therefore modifies the relative weights of the different partonic channels across the measured phase space, and consequently the quark--gluon composition of the reconstructed jet sample.
For the forward kinematics considered here, increasing the center-of-mass energy from $\sqrt{s}=8~\mathrm{TeV}$ to 13 TeV leads to a more quark-enriched sample at fixed $p_{JT}$.
The resulting EEC distribution is therefore more strongly dominated by the quark-initiated contribution, providing even cleaner access to the non-perturbative confinement transition of quark jets.

\bibliography{refs.bib}
\bibliographystyle{JHEP-2modlong.bst}

\end{document}